\documentclass[10pt,conference]{IEEEtran}

\makeatletter
 \def\ps@headings{%
 \def\@oddhead{\mbox{}\scriptsize\rightmark \hfil \thepage}%
 \def\@evenhead{\scriptsize\thepage \hfil \leftmark\mbox{}}%
 \def\@oddfoot{}%
 \def\@evenfoot{}}
 \makeatother
\IEEEoverridecommandlockouts

\ifx\pdfoutput\undefined
\usepackage{graphicx}
\else
\usepackage[pdftex,final]{graphicx}
\fi

\usepackage{subfigure}
\usepackage{cite}
\usepackage{epsfig,amsmath,amssymb,graphicx}
\usepackage{pdfsync}
\usepackage{fancyhdr}
\usepackage{lastpage}
\usepackage{color}
\usepackage{times,rawfonts}

\usepackage[draft]{hyperref}
\usepackage{algorithm2e}
\usepackage{multicol}

\fancypagestyle{titlepagestyle}{
 \fancyhf{}
}

\renewcommand{\headrulewidth}{0pt}
\begin{document}

%

    \renewcommand{\sectionautorefname}{Section}
    \renewcommand{\figureautorefname}{Fig.}

\renewcommand{\L}{\ensuremath{\operatorname{L}}}
\newcommand{\AM}{\ensuremath{\mathcal{A}_M}}
\newcommand{\PM}{\ensuremath{\mathcal{P}_M}}

\title{\Large \bf OPTIMIZING NETWORK TOPOLOGY TO REDUCE AGGREGATE TRAFFIC
IN SYSTEMS OF MOBILE ROBOTS}
\author{
    Leenhapat Navaravong, John M.\ Shea \\
    University of Florida,\\
    Gainesville, FL \\
    \texttt{{\small\{leenhapat@ufl.edu,
        jshea@ece.ufl.edu\}}}
  \and
    Eduardo L.\ Pasiliao Jr, Gregory L.\ Barnette \\
    Eglin AFB, FL \\
    \texttt{{\small\{pasiliao@eglin.af.mil\}}}\\
    \texttt{{\small\{gregory.barnette@eglin.af.mil\}}}
  \and
    Warren E.\ Dixon \\
    University of Florida,\\
    Gainesville, FL \\
\texttt{{\small\{wdixon@ufl.edu\}}}
    }
\date{}
\maketitle


\begin{abstract}
  Systems of networked mobile robots, such as unmanned aerial or
  ground vehicles, will play important roles in future military and
  commercial applications.  The communications for such systems will
  typically be over wireless links and may require that the robots
  form an ad hoc network and communicate on a peer-to-peer basis. In
  this paper, we consider the problem of optimizing the network
  topology to minimize the total traffic in a network required to
  support a given set of data flows under constraints on the amount of
  movement possible at each mobile robot.  In this paper, we consider
  a subclass of this problem in which the initial and final topologies
  are trees, and the movement restrictions are given in terms of
  the number of edges in the graph that must be traversed.  We develop
  algorithms to optimize the network topology while maintaining
  network connectivity during the topology reconfiguration process.
  Our topology reconfiguration algorithm uses the concept of prefix
  labelling and routing to move nodes through the network while
  maintaining network connectivity.  We develop two algorithms to
  determine the final network topology: an optimal, but
  computationally complex algorithm, and a greedy suboptimal algorithm
  that has much lower complexity.  We present simulation results to
  compare the performance of these algorithm.
\end{abstract}

\section{Introduction}\label{sec:intro}

Autonomous unmanned aerial or ground vehicles, which function as
systems of networked mobile robots, will play important roles in
future military and commercial applications. The communications for
such systems will typically be over wireless links and may require
that the robots form an ad hoc network and communicate on a
peer-to-peer basis~\cite{Freeb01,Giu00,Vcom93}. In this scenario, the
total amount of traffic generated in sending information across the
network will depend both on the information flows to be transmitted,
as well as the topology of the network. The latter consideration is
because of the need for intermediate nodes to relay information
between a source and destination.  Thus, the {\it aggregate data
  traffic}, which includes all of the data transmissions from sources
and relays will generally be much larger than the total traffic flow
from the sources.  In this paper, we focus on data traffic only and do
not consider the impact of control traffic, and thus use the term {\it
  aggregate traffic} in place of aggregate data traffic from here on.

Since the robots are mobile, the aggregate traffic can be reduced by
reconfiguring the network topology to move some of the communicating
robots closer together.  We consider networks in which the network
connectivity must be maintained at all times, and any movement scheme
must take this into account.  In addition, the mobile robots may have
finite energy that limit the extent of their movement or may be
otherwise constrained in their movement because of their other duties,
such as sensing.  Thus, we consider the problem of optimizing the
network topology to minimize the aggregate traffic in a network to
support a given set of data flows, under constraints on the amount of
movement possible at each mobile robot.  In the case that the mobile
robots do not have any energy constraints and the shape of the final
network topology (a graph consisting of sets of edges and vertices,
but not the assignment of robots to vertices) is already defined, this
problem falls in the class of resource allocation problems known as
quadratic assignment problems~\cite{CW05,Garey}.  Unfortunately, even
for this simpler subclass of problems, the problem is NP-Hard, and
thus there are no known solutions that run in polynomial time.

There are many previous papers on formation control of mobile robots.
For instance~\cite{Herbert05} considers centralized solutions to
reconfigure the physical topology of a group of networked mobile
robots to achieve a desired final topology while avoiding obstacles
and collision.  In~\cite{Maria06}, a decentralized topology control
approach is presented, but network connectivity is not considered.
In~\cite{ZhenTAC11,KanMILCOM10}, a decentralized topology control
approach is developed to achieve a desired physical network formation
while maintaining network connectivity, given that the network is
already in the desired network topology.  In~\cite{BomMILCOM11}, new
approaches are developed to reconfigure a network topology from an
arbitrary initial connected graph to a specified desired tree
topology, when there are no constraints on the amount of movement of
the nodes.  The fundamental idea of the approach in~\cite{BomMILCOM11}
is that robots that are not in the desired topology are ``routed''
through the network topology to transform the network while
maintaining connectivity.  In~\cite{BomMILCOM11}, all nodes are
considered identical, and prefix labelling and routing techniques
(cf.~\cite{JJGarciaICCCN09,Sampath09,JJGarciaMASS09}) are used to
assign labels and routes.

In this paper, we consider problems where the initial topology is
given, but the final topology must be chosen to minimize the aggregate
traffic in the network, under constraints on the amount of movement of
the robots.  We consider the scenario in which the initial and final
network topologies are trees\footnote{Any connect graph always has a
  connected tree subtopology -- a spanning tree}.  In our optimization
algorithms, we use the amount of movement required for topology
reconfiguration with a prefix routing approach, so that network
connectivity is ensured at all times.  We find exact and greedy
algorithms to minimize the aggregate traffic.  The performance of the
algorithms are evaluated and compared using simulations.


\section{Problem Formulation}
\label{sec:form}

We consider a system of mobile robots that communicate over wireless
links with limited communication distance.  It is convenient to
represent the induced network topology as a simple graph $\mathcal{G} =
(\mathcal{V},\mathcal{E})$, where vertices $\mathcal{V}$ represent the
robots, and an edge $e \in \mathcal{E}$ between vertices $u$ and $v$
indicates that $u$ and $v$ can communicate over a wireless link.  Let
$\mathcal{F}=\left\{ f_{(u,v)}: ~ (u,v) \in \mathcal{G}^2 \right\}$ be
the set of data flows, where $f_{(u,v)}$ denote the amount of traffic
from source $u$ to destination $v$ and $\mathcal{G}^2$ denotes the
Cartesian product of $\mathcal{G}$ with itself.  Then the aggregate
traffic over network topology $\mathcal{G}$ is
\begin{equation}
\label{eq:initialcost}\sum_{(u,v) \in
  \mathcal{G}^2}f_{(u,v)}d_{\mathcal{G}}(u,v),
\end{equation}
where the distance function $d_{\mathcal{G}}(u,v)$ is the number of
edges in the shortest path between vertices $u$ and $v$ in
$\mathcal{G}$.  The initial network topology is assumed to be a tree
and is labeled $\mathcal{G}_{i}$.

In order to facilitate our later algorithms, a root node is chosen in
$\mathcal{G}_i$, and prefix labeling is applied starting at the root
to give a prefix tree, or {\it trie}.  The distance between nodes in
the trie can be simply determined by its prefix label. We first find
the largest prefix that is common to the labels of both nodes.  This
is the prefix label of a common parent of both node in the tree.  Then
the shortest path between two nodes is up to the common parent and
then back down to the other node. Hence the total distance is sum of
the distance from each of them to their common parent.  Let
$\Lambda_u$ denote the prefix label assigned to node $u$.  Let
$\L(\Lambda_u)$ denote the length of the prefix label of node $u$, and
$\L(\Lambda_u,\Lambda_v)$ denote the maximum length prefix in common
to the prefix labels of nodes $u$ and $v$.  Then
\begin{equation}
\label{eq:distance_1} d_{\mathcal{G}}(u,v) = [\L(\Lambda_{u}) -
\L(\Lambda_{u},\Lambda_{v})] + [\L(\Lambda_{v}) -
\L(\Lambda_{u},\Lambda_{v})].
\end{equation}

As can be seen from \eqref{eq:initialcost}, the larger the distance
between two nodes that share a data flow, the greater the aggregate
traffic in the network, since the same message will be relayed at
every intermediate node between them.  To minimize the aggregate
traffic in the absence of any energy constraints, then any final connected graph
topology $\mathcal{G}_f$ is possible.  Let $\mathbb{C}(\mathcal{G})$
be the connectivity function, which takes on the value 1 when the
final topology is connected and 0 otherwise.  Then
we wish to find $\mathcal{G}_{f}$ that satisfies
\begin{align}
\label{eq:min} 
 \mathcal{G}_f = \arg & \min_{\mathcal{G}}
\sum_{(u,v) \in \mathcal{G}^2} f_{uv} d_{\mathcal{G}}(u,v)\\
& \mbox{subject to} \nonumber \\
& \mathbb{C}(\mathcal{G})=1. \nonumber
\end{align}

Now, if we constrain that each node has limited energy, then some
final graph topologies may no longer be possible.  Moreover, the
constraint that the network be connected at all times will also limit
which final topologies are possible in this scenario.  For instance,
if a node is to move up the tree, then all of its children must have
sufficient energy to at least move up to connect with that node's
parent.  Let $h_u$ denote the number of vertices in the graph that
each node may move before its energy budget is expended, and let
$\mathcal{H}= \{ h_u : u \in \mathbf{G} \}$. Let
$\mathbb{F}(\mathcal{G}_{i},\mathcal{G}_{f},\mathcal{H})$ be a
feasibility function, which takes on the value 1 when the final
topology is feasible under the energy constraint and 0 otherwise.
Then the aggregate traffic minimization under the energy and network
connectivity constraints can be formulated as:
\begin{align}
\label{eq:min_energy_constraint} 
 \mathcal{G}_f = \arg & \min_{\mathcal{G}}
\sum_{(u,v) \in \mathcal{G}^2} f_{uv} d_{\mathcal{G}}(u,v)\\
& \mbox{subject to} \nonumber \\
& \mathbb{C}(\mathcal{G})=1 \nonumber \\
& \mathbb{F}(\mathcal{G}_{i},\mathcal{G},\mathcal{H})=1. \nonumber
\end{align}
We determine $ \mathbb{F}(\mathcal{G}_i,\mathcal{G},\mathcal{H})$
based on transforming the topology using the prefix-routing approach
described in the next section.


\section{Network Topology Configuration Algorithms}
\label{sec:network}
\begin{figure}[t]
\centering
\subfigure[Initial network topology.] 
{
    \label{fig:initial_1}
    \includegraphics[scale=0.49]{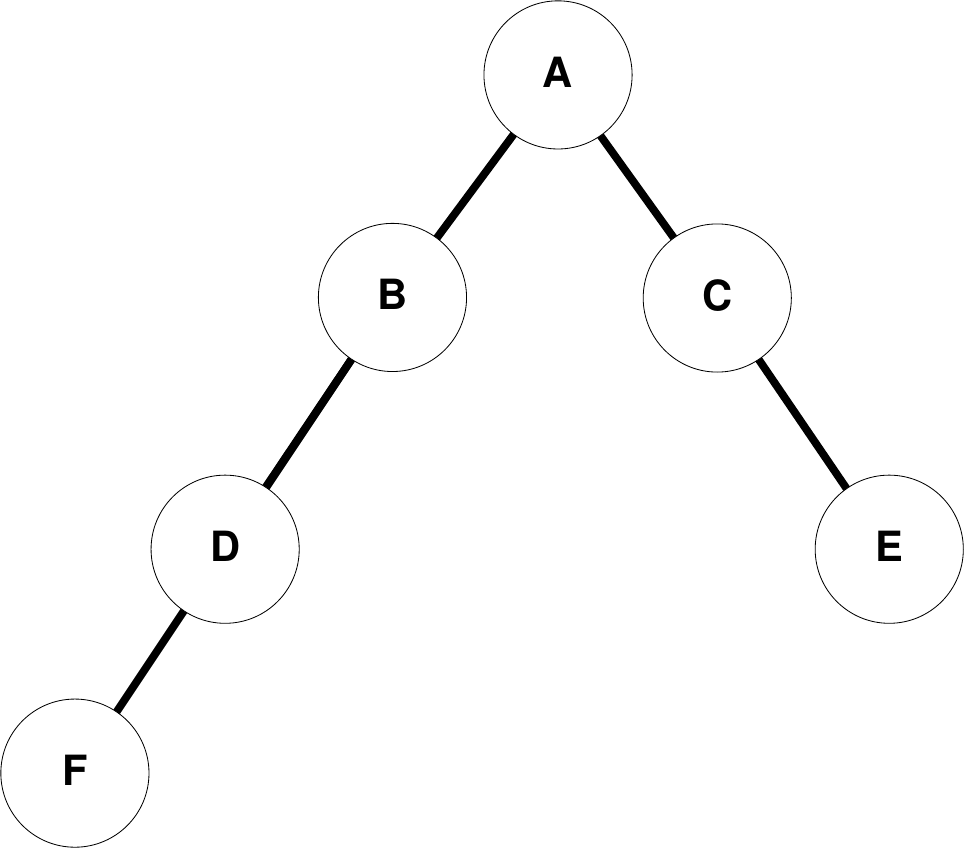}
  }
\subfigure[Desired network topology.] 
  {
    \label{fig:desired_1}
    \includegraphics[scale=0.49]{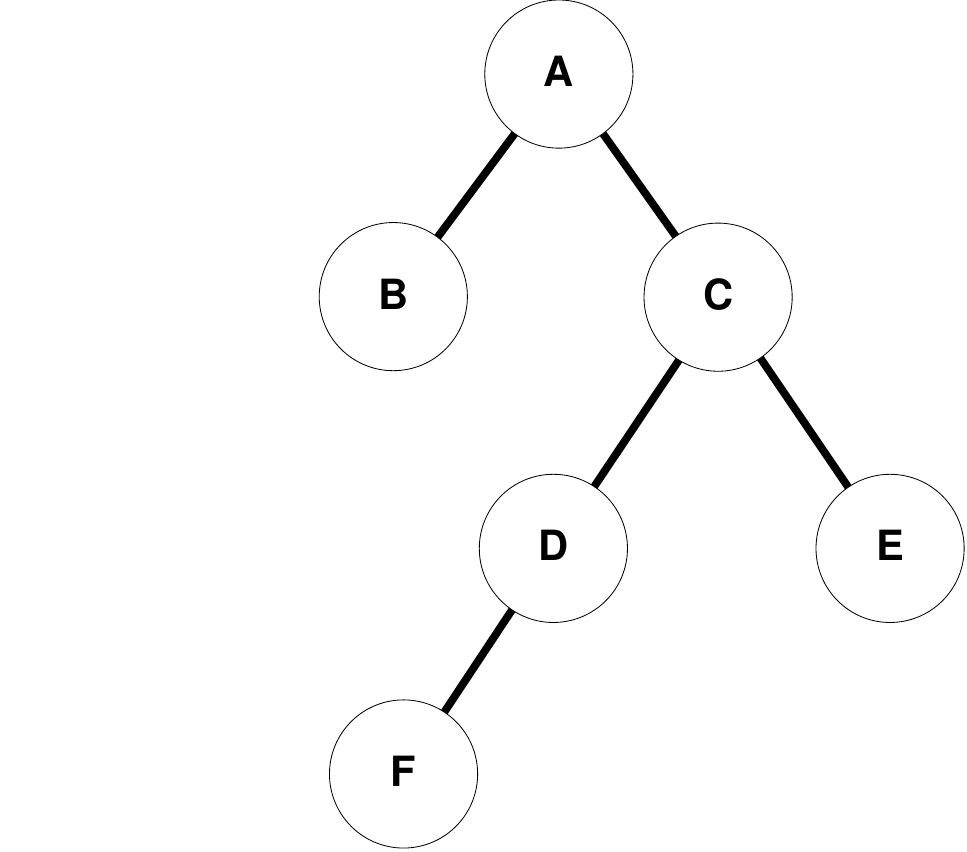}
}\caption{Network topology.}
\label{fig:network_1}
\end{figure}

Before addressing techniques to solve
\eqref{eq:min_energy_constraint}, we describe how the network topology
control method of~\cite{BomMILCOM11} can be utilized in this
application, in which nodes are not identical.  In this section, we
assume that both the initial topology $\mathcal{G}_i$ and the final
topology $\mathcal{G}_f$ are known.  We begin by choosing a node in
the initial topology to serve as the root of the tree.  In this paper,
we assume that the root is chosen at random.  As an alternative, the
root may also be chosen according to some criteria; the design of
root-selection algorithms is outside the scope of this paper.  As
mentioned in \autoref{sec:form}, the root then assigns unique prefix
labels to each of its children, which assign unique prefix labels to
their children, etc., until the entire tree has prefix labels.  In
prefix labels, the label of a vertex's parent node is a prefix of that
node's label.  The initial tree topology $\mathcal{G}_{i}$ becomes a
prefix tree, or {\it trie}~\cite{CLRbook,Drozdek,Horowitz}.  The
prefix label assigned to each node serves as its network address.

\begin{figure}[t]
\centering
\subfigure[Initial network topology.] 
{
    \label{fig:initial_2}
    \includegraphics[scale=0.49]{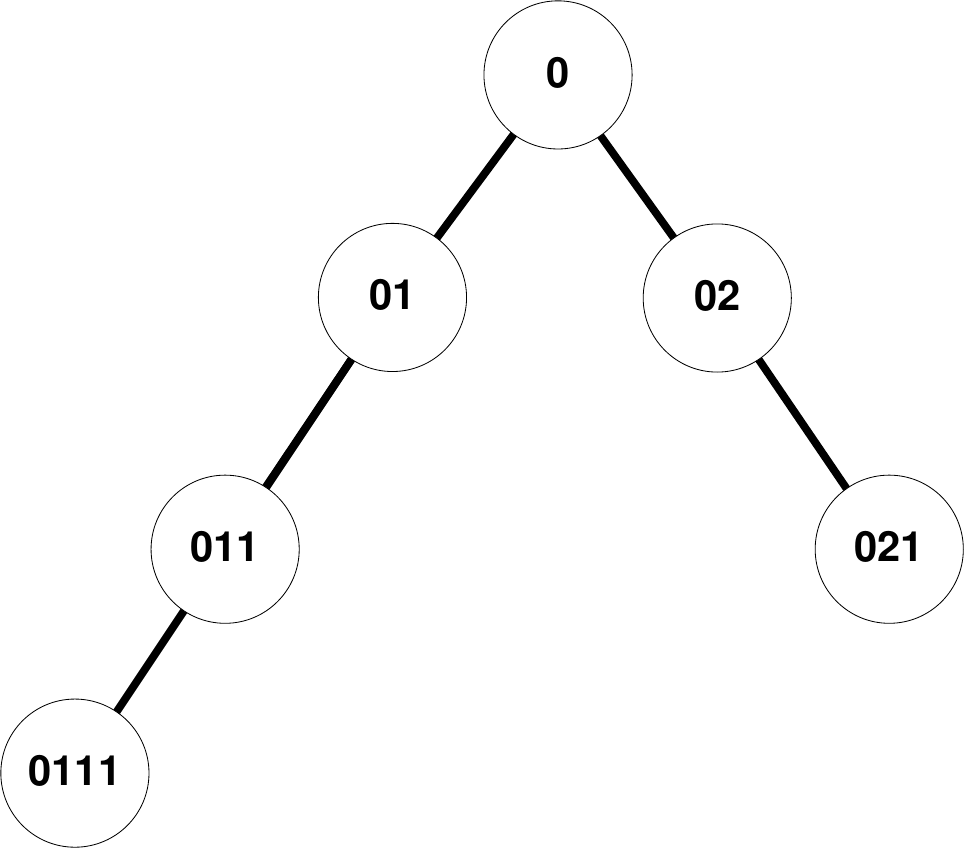}
  }
\subfigure[Desired network topology.] 
  {
    \label{fig:desired_2}
    \includegraphics[scale=0.49]{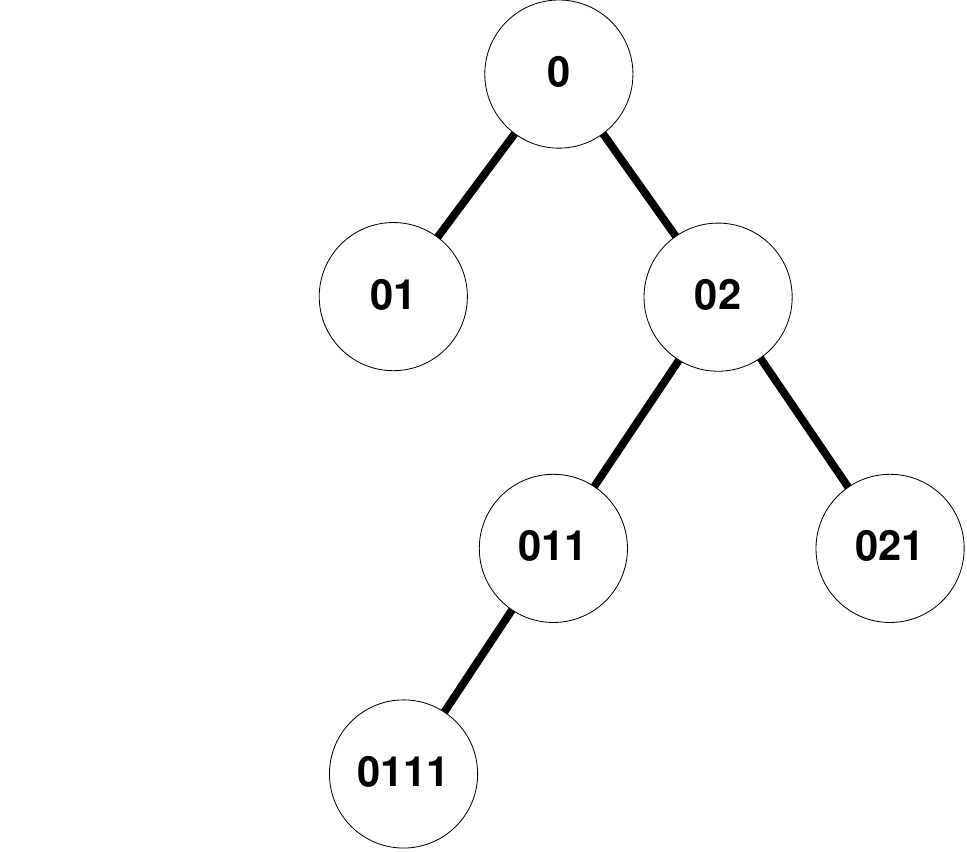}
}\caption{Labeling network topology.} \label{fig:network_2}
\end{figure}

We explain the prefix-routing approach to network topology control
using the example topologies shown in \autoref{fig:network_1}.  Node~A
has been selected to be the root.  Prefix labels are then assigned to
all nodes in the initial network starting from a root, as shown in
\autoref{fig:initial_2}.  After prefix label assignment is done for
the initial network tree, each node sends a message including its own
prefix label and identity to the root. After the root obtains all
messages from each node, it will have a knowledge of the initial
network graph.  The root will then label all the nodes in the desired
network tree with the prefix label assigned to the same node in the
initial tree.  The desired network tree after label assignment is
completed is shown in~\autoref{fig:desired_2}.

The root searches for nodes that need to move between the initial and
final topologies, starting from the top to the bottom of the tree, in
a breadth-first manner.  The nodes that must move are those whose
prefix label does not correspond to the position where it is located
in the desired tree.  The label for a node should always be of the
form
\begin{equation}
\label{eq:prefix}
 \Lambda = \Lambda_{parent}\odot l,
\end{equation}
where $\Lambda$ denote the prefix label of parent's children, $\odot$
is the concatenate operator, and $l$ is the unique suffix.  Nodes that
do not have the correct prefix label must move from their position in
the initial topology, and hence are called {\it moving nodes}.  Nodes
that have the correct prefix label and that have not been previously
assigned to be a moving node (see more below) are {\it nonmoving nodes}.

For the example network, all of the nodes that are one edge away
from the root have the correct prefix label and thus are nonmoving
nodes. Next, the root considers all nodes that are two edges away
(it's children's descendants).  As shown in \autoref{fig:desired_2},
the node with label $021$ has a correct prefix label, but the node
with label $011$ does not have a correct prefix label.  Thus, node
$011$ will be a moving node.  If a parent moves, it will cause
network connectivity to break for its children, so all of the
descendants of a moving node must also be moving nodes.  For instance,
since $011$ is a moving node, its child $0111$ must also be a moving node.
So, even though $0111$ initially has a prefix label that matches its
parent in \autoref{fig:desired_2}, it is still a moving node.

For each moving node, the root records two labels: (1)
its \begin{it}anchor-node label \end{it} is the label of the
non-moving node that will be the moving node's destination, and (2)
its \begin{it} desired label\end{it} is the new label of the moving
node upon an arrival at the destination in the desired topology.  When
the root has already considered all nodes in the desired tree, the
root will send a message {\bf M.Dest} including both labels to each
moving node. A moving node then first move to the node whose prefix
label is the anchor-node label.  When a moving node arrives at a
non-moving node, the non-moving node first looks for the moving node's
anchor-node label to see if it match its prefix label. If it does, it
will serve as the anchor node for that moving node, and it then uses
the desired label of that moving node to forward the moving node to
the right position in the desired graph.  The moving node will be
relabeled to match the desired label once it reaches its final
position, which will make its prefix label correspond to
its position in the desired network topology.

The desired label of a moving node can be simply
determined from its parent in the desired topology as given
in~\autoref{fig:desired_2}. If its parent is a moving node, its
parent must already be assigned the desired label by the root, and the
moving node's desired label is determined from  the desired label
of its parent. If its parent is a nonmoving node, the desired
label is determined from its parent prefix label.

When the moving nodes 011 and 0111 receive a message {\bf M.Dest} from
the root including both the anchor-node and desired labels, it will
move through the initial network toward anchor node 02 by using
maximum prefix matching logic.  When a moving node 011 and 0111 are
able to connect to the anchor node 02, anchor node 02 will look at
their anchor labels to check if 02 is their anchor node.  Once node 02
determines that it is the anchor node for 011 and 0111, node 02 will
check the desired label of both of the nodes.  The desired labels are
022 and 0221, respectively, and 02 will use these labels to forward
nodes 011 and 0111 to the right positions in the desired topology.
After both nodes arrive at the desired position, their labels will be
changed to the desired labels, which will make their prefix labels
correspond to their position in the desired network topology.

\begin{figure*}[t]
  \centering \subfigure[The root node, 0, sends message {\bf M.Dest}
  including the anchor-node and desired labels to the moving nodes, 011
  and 0111.] 
  {
    \label{fig:trans_1}
    \includegraphics[scale=0.68]{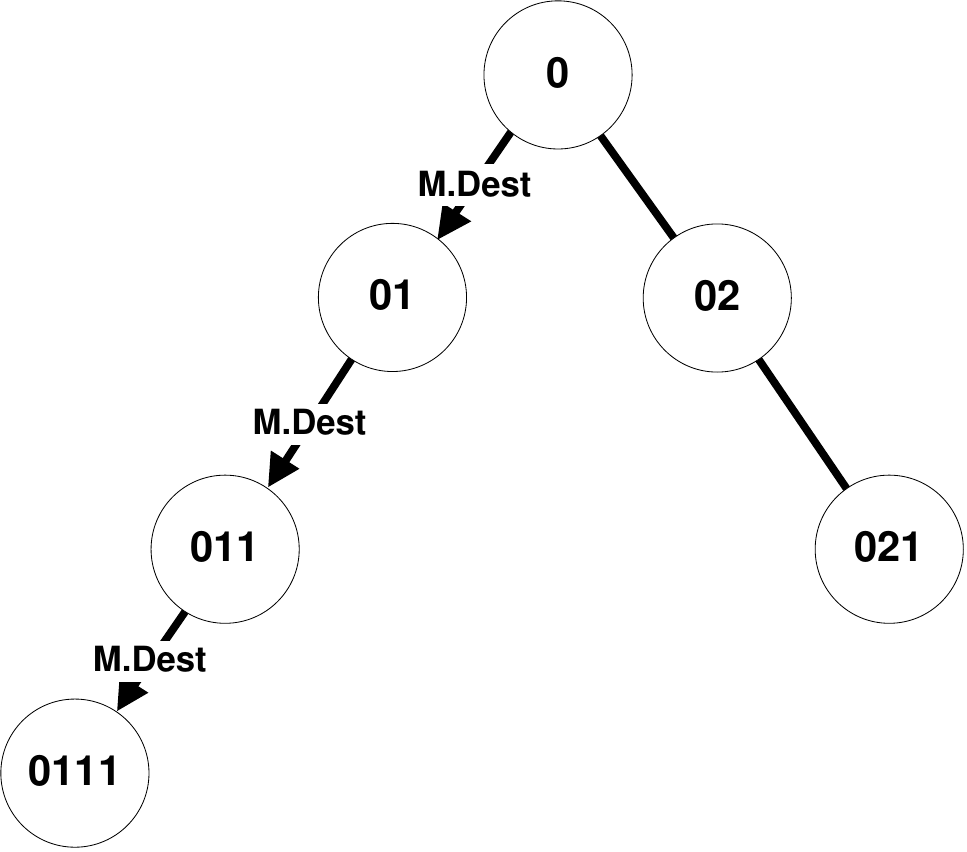}
  } \subfigure[Nodes 011 and 0111 use prefix routing to move through
  the tree until they are able to connect to their designated anchor
  node, 02.] 
  {
    \label{fig:trans_2}
    \includegraphics[scale=0.68]{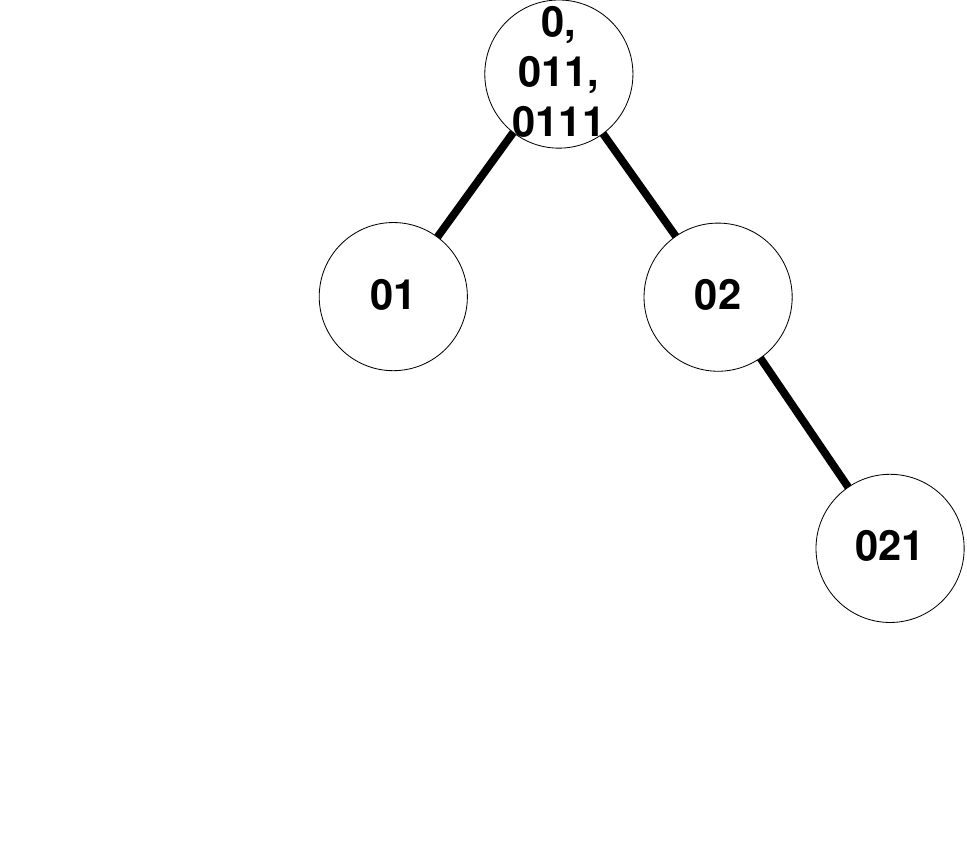}
  } \subfigure[Node 011 is forward to an appropriate position in the
  desired topology and relabeled as 022.] 
  {
    \label{fig:trans_3}
    \includegraphics[scale=0.68]{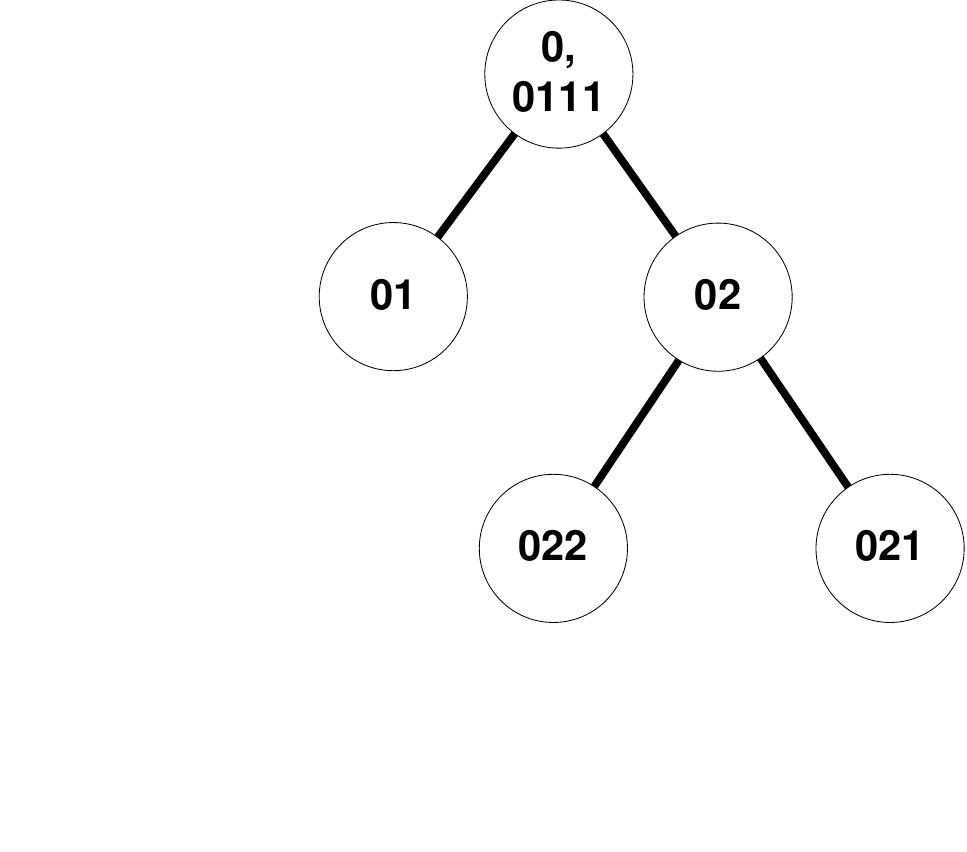}
  } \subfigure[Node 0111 moves toward the anchor node 02, which allows
  it to connect to node 022.] 
  {
    \label{fig:trans_4}
    \includegraphics[scale=0.68]{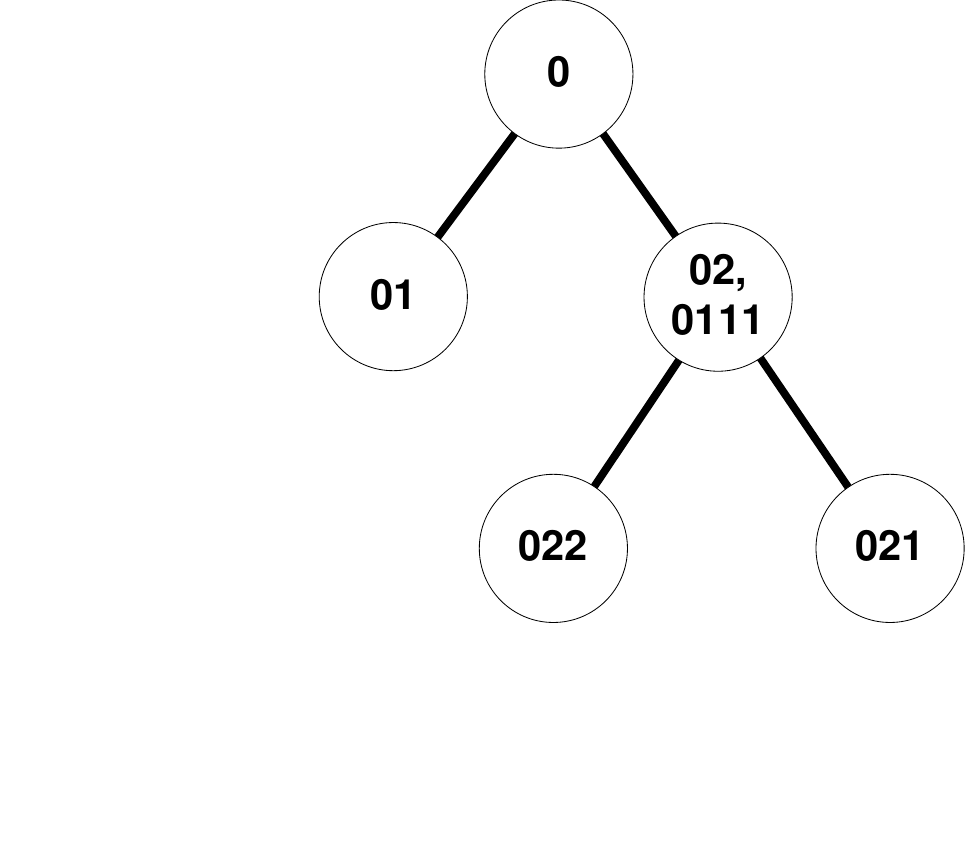}
  } \subfigure[Node 0111 is relabeled as 0221 to achieve the desired prefix
  tree topology.] 
  {
    \label{fig:trans_5}
    \includegraphics[scale=0.68]{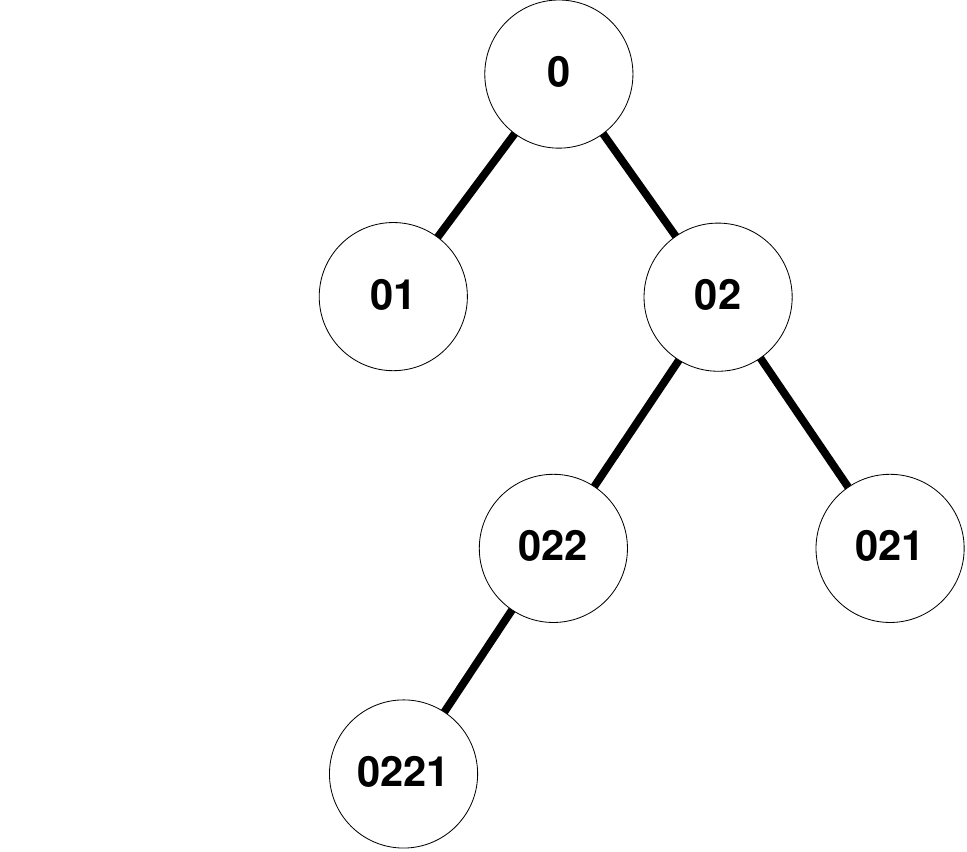}
}\caption{Network reconfiguration.} \label{fig:network_3}
\end{figure*}

Consider now how nodes should be move from their positions in the
initial topology to their positions in the desired topology without
breaking network connectivity.  Generally, if there are multiple
moving nodes in the initial topology, whenever they receive a message
from a root, they can start moving simultaneously. However, a moving
node that is not a leaf node has to wait for its descendants to move
up to it before it can start moving.  Otherwise, the node's
descendants will be disconnected from the network.  For example,
consider again nodes 011 and 0111 in \autoref{fig:initial_2}. Node 011
cannot move first, since that would break network connectivity to node
0111; in general, a parent node cannot move -- all of its children
must move first to make it a leaf node.  Thus, node 0111 first has to
move up to node 011 until it is able to connect to node 01.  Then both
nodes 011 and 0111 can continue moving up to the root, passing by node
01, until they are able to connect with node 02.  Once node 011
connects with 02 (at the time it reaches the root), it will be
immediately relabeled as 022 to make the label of the node 011 conform
the prefix tree.  Then node 0111 will move toward the node 02 until it
is able to connect to node 011 which is already relabeled as 022
. Finally, the node 0111 will be relabeled as 0221 to achieve the
desired topology that has all the node's label conform the prefix
tree.  An example of this method is shown in~\autoref{fig:network_3},
and the pseudocode of a label assignment algorithm for a root is given
in Algorithm 1.

\begin{algorithm}[t]

\begin{tabular}[c]{p{8cm}}

\hline
{\scriptsize \begin{bf}Algorithm 1: \end{bf} Label Assignment }\\
\hline
\SetKwFunction{GC}{GetChildren}
\SetKwFunction{GR}{GetRoot}
\SetKwFunction{Label}{Label}
\SetKwFunction{PL}{ParentLabel}
\SetKwFunction{GDL}{GetUniqueDesiredLabel}

{\scriptsize \KwIn{$\mathcal{G}_{f} = \{\mathcal{V},\mathcal{E}_{f}\}$ } \tcc{desired graph with corresponding prefix label from
$\mathcal{G}_{i}$}

\KwOut{$\mathcal{G}_{f}$  with moving indicators and anchor-node and
  desired labels}

$root$=\GR{$\mathcal{V}$}\;
$root.moving$ = {\bf false}\;
$Q= \{ root \}$\;
$Qtmp = \emptyset$\;
$ConsideredNodes = Q$\;

\While{$|ConsideredNodes| < |\mathcal{V}|$}{

\ForEach{v$\in$Q}{

$\mathcal{C}$ = \GC{v}\;
\If{$\mathcal{C} \neq \emptyset$ }
{
$Qtmp = Qtmp \cup \mathcal{C}$\;
$ConsideredNodes = ConsideredNodes \cup  \mathcal{C}$\;

\eIf{ v.moving == {\bf false} }
{
\ForEach{$c \in \mathcal{C}$}
{
\eIf{\PL{c}== \Label{v}}
{
$c.moving$ = {\bf false}\;
}
{
$c.moving$ = {\bf true}\;
}
}
\ForEach{$c \in \mathcal{C}$}
{
\If{$c.moving$ == {\bf true}}
{
$c.anchorlabel = v.label$\;
$c.desiredlabel$ = \GDL{}\;}}}
{
\ForEach{$c \in \mathcal{C}$}{

$c.moving$ = {\bf true}\;
$c.anchorlabel = v.anchorlabel$\;
$c.desiredlabel$ = \GDL{}\;

}

}

}

}

Q = Qtmp\;

}

}
\\
\hline

\end{tabular}

\end{algorithm}

\section{Network Topology Optimization Algorithms}
\label{sec:opt}

In this section, we present techniques to solve
\eqref{eq:min_energy_constraint} under the additional constraint that
the network topology is reconfigured around a root node, as described
in \autoref{sec:network}.  The root controls the network topology and
wishes to select a final topology that minimizes the aggregate
traffic, under the energy constraints that limit each node's movement.
The distance that a node must move to transition from the initial to
the final topology is
\begin{equation}
\label{eq:distance_2} \underset{\mathcal{G}_{i} \rightarrow \mathcal{G}_{f}}{d}(v) = d_{\mathcal{G}_{i}}(v,a_{v}) +
d_{\mathcal{G}_{f}}(v,a_{v}) - 2,
\end{equation}
where $v$ is the moving node and $a_{v}$ is the anchor node of $v$.
For instance in the example of \autoref{fig:network_2}, node $D$ must move
to within one hop of its new parent, which is $C$ or $02$.  Thus, node
$D$ moves up to the root, at which point it is within one hop of $02$
and can thus be relabeled $022$ to achieve the desired position by only
moving two hops.

With the additional constraint that the topology reconfiguration occurs
around the root and using the constraints on the amount of movement of
a node, the optimization problem can be formulated as
\begin{align}
\label{eq:rooted_optimization} 
 \mathcal{G}_f = \arg & \min_{\mathcal{G}}
\sum_{(u,v) \in \mathcal{G}^2} f_{uv} d_{\mathcal{G}}(u,v)\\
 \mbox{subject to} \nonumber\\
&\mathbb{C}(\mathcal{G})=1 \nonumber \\
& \underset{\mathcal{G}_{i} \rightarrow
\mathcal{G}_{f}}{d}(v) \leq h_{v}, \forall v \in (\mathcal{V}\setminus
v_{\text{root}}) \nonumber \\
& \underset{\mathcal{G}_{i} \rightarrow
  \mathcal{G}_{f}}{d}(v_{\text{root}}) = 0\nonumber
\end{align}

Before presenting algorithms to solve this problem, we first consider
the necessary scope of the search by evaluating which nodes may need
be moved between $\mathcal{G}_i$ and $\mathcal{G}_f$.  We partition
the nodes in to {\it active nodes}, which have a data flow to or from
other nodes, and {\it passive nodes}, which do not have a data flow to
or from other nodes.  Note that passive nodes may still act as relays
for other nodes' data flows.  To conserve energy, it is best to not
move passive nodes unless it is required to allow active nodes to
move.  To decide which node should be repositioned, we first
considered those active nodes that have enough energy to move at least
one hop.  Such nodes are the initial members of the {\it active moving
  node set}, $\mathcal{A}_M$.  However, the initial members of
$\mathcal{A}_M$ may not all be free to move because network
connectivity must be maintained, and under our prefix topology
reconfiguration approach, a node cannot move while it still has
children.
Thus, for a node to remain in $\mathcal{A}_M$, all of it
descendants must have sufficient energy to reach that node and hence
be able to establish communication with that node's parent.
Hence all descendants of a node in $\mathcal{A}_M$ must be able to
move a number of hops given by
\begin{equation}
\label{eq:hop}
 h_{v_{c}} \geq  d_{\mathcal{G}_{i}}(v,v_{d}),
\end{equation}
where $v_{d}$ denote the descendant of node $v$ belonging to
$\mathcal{A}_M$.
Any of the descendants of a moving node $v$ must have enough energy to
move up at least on hop to properly connect to the parent node of $v$
to maintain the network connectivity.  Nodes in $\mathcal{A}_M$ that
do not have any children are free to move as far as their energy
constraint allows.  The nodes in $\mathcal{A}_M$ whose descendants'
maximum possible movements $h_{v}$ do not satisfy~\autoref{eq:hop}
are removed from $\mathcal{A}_M$ because they cannot be moved.
Furthermore, any passive nodes that are children of nodes that remain
in $\mathcal{A}_M$ may be moved and are put in another set called the
{\it passive moving node set}, $\mathcal{P}_M$.  This node
classification algorithm is formalized in Algorithm 2.


\begin{algorithm}[htbp]

\begin{tabular}[c]{p{8cm}}

\hline
{\scriptsize \begin{bf}Algorithm 2: \end{bf} Moving Node Selection }\\
\hline
\SetKwFunction{GC}{GetChildren}
\SetKwFunction{CA}{CheckFlow}
\SetKwFunction{RP}{ReachParentPossible}

{\scriptsize \KwIn{The initial graph, $\mathcal{G}_{i} = \{\mathcal{V},\mathcal{E}_{i}\}$ }
{\scriptsize \KwOut{The active moving set, $\mathcal{A}_M$ and the
    passive moving set, $\mathcal{P}_M$}

\CA(v) $\overset{\Delta}{=} \bigcup_{u \in \mathcal{V} \backslash v} \left( f_{uv} >0 \cup f_{vu} >0 \right)$;

\RP{v,$\mathcal{C}$} $\overset{\Delta}{=} \bigcup_{c \in \mathcal{C}}
\left( h_{c} \geq  d_{\mathcal{G}_{i}}(v,c) \right)$ \;

\ForEach{$v \in \mathcal{V}$}{

\If{($\operatorname{CheckFlow}(v)$ == {\bf true}) $\&\&$ ($h_{v} \geq 1$) }{

$\mathcal{C} =$ \GC{v}\;
\eIf{$\mathcal{C} == \emptyset$}{ $\mathcal{A}_M = \mathcal{A}_M
  \cup v$\; }{
\If{\RP{v,$\mathcal{C}$} == {\bf true} } { $\mathcal{A}_M =
  \mathcal{A}_M \cup v$\;
\ForEach{c $\in$ C}{ \If{\CA{c} == {\bf
false}} { $\mathcal{P}_M = \mathcal{P}_M \cup c$\; } }

}

}

}

}

}

}

\\
\hline

\end{tabular}

\end{algorithm}

After the root finds \AM\ and \PM , a subgraph $\mathcal{G}' \subset
\mathcal{G}_{i}$ is formed by removing all vertices in \AM\ and \PM,
along with all associated edges.  All the nodes from both sets can move
and will become descendants of at least one node in $\mathcal{G}'$
according to the optimal or greedy algorithms described below.

\subsection{Optimal Algorithm}
\label{sec:ea}

For a root to achieve an optimal achievable topology, it essentially
has to consider all possible tree topologies and select the final
topology based on the achievable tree that gives the minimum aggregate
data traffic.  In this section, we provide details about how the
the optimal solution can be found, subject to the constraint that
the topology is reorganized around a pre-selected root.  We use the
branch-and-bound technique to limit the complexity of this
combinatorial search.

As described above, the root first obtains the active and passive
moving node sets using Algorithm 2, as well as the subgraph
$\mathcal{G}'$ of non-moving nodes.  A brute-force solution is for the
root to consider all nodes in the active moving node set and to
evaluate the aggregate data for all achievable tree topologies
$\mathcal{G}$ such that $ \mathcal{G}' \subset \mathcal{G}
:(\forall{\mathcal{G}})$.  This can be done by finding all
permutations of the nodes in \AM\ and then for each permutation, each
node from this permutation is attached in order one by one from the
first to the last to a tree in every possible way.  The order of the
placement is important because the nodes in \AM\ can attach not only
to the nodes in $\mathcal{G}'$ but also to other nodes in \AM\
that have already been placed.  After the nodes in \AM\ are placed,
the nodes that need to move in \PM\ are then repositioned to the
places that are closest to their original positions.


The complexity of the brute-force combinatorial search can be reduced
by applying the branch-and-bound method~\cite{land60a}.  The idea in
branch-and-bound is that all partial and complete solutions are
represented by nodes on a tree, in which a leaf of this tree indicates
a complete solution.  The search for the best solution starts from the
root of the tree.  At each search node, the algorithm tries to
determine if a branch can be pruned, which is possible if the lower
bound on the aggregate traffic is greater than the upper bound for the
aggregate traffic in some other branch, as such branches can never
yield the optimum solution.  The search is performed until all the
nodes in the tree are examined or pruned.

In this paper, we use a simple approach to branch-and-bound based on a
depth-first tree search across node assignments, one permutation at a
time.  For convenience of description, we index the levels of the
tree, where the root is defined to be at level -1.  The children of
the root are at level 0 and represent all possible permutations of \AM
.  At level~1 are all possible locations for the first node in the
permutation of \AM .  At level~$n$ are all possible locations for the
$n$th node in the permutation of \AM\ given all the previous locations
of nodes $1,2,\ldots,n-1$, which are determined by the $n$th node's
parent.  The leaves represent a complete solution
$\mathcal{G}_{complete}$ for a particular permutation of \AM .

The search proceeds in depth-first fashion, first by selecting one
permutation and then by trying one allocation of all nodes.  At each
node the minimum aggregate traffic can be lower bounded by the
aggregate traffic from the nodes that have already been assigned
positions plus the sum of the remaining data flows.  Once the
depth-first search has reached a leaf node, we have one possible
solution to the minimum aggregate traffic, and we use this as an upper
bound on the best minimum aggregate traffic over all nodes.  Then as
we proceed down other branches, we eliminate a branch whenever the
lower bound for that branch exceeds the upper bound on the optimal
solution, which is given by the best solution found so far.  Whenever
the search reaches a leaf of the tree, the aggregate traffic will be
checked and compared with the best solution found. If this value is
better than the best solution, it will then be recorded as a new best
feasible complete solution and this complete solution
$\mathcal{G}_{complete}$ will also be recorded as the best possible
solution found. The optimal solution is found when all nodes
have been considered or pruned.  Because our branch-and-bound approach
uses depth-first search, it is most easily implemented using
recursion, and we omit the detailed algorithm here.  To give an idea
of the working of this algorithm, we give a nonrecursive form of the
algorithm for an active moving node set with three nodes in
Algorithm~3.

\begin{algorithm}[t]

\begin{tabular}[c]{p{8cm}}

\hline
{\scriptsize \begin{bf}Algorithm 3: \end{bf} Optimal Algorithm with
  Branch-and-Bound for an Active Moving Node Set with 3 Nodes}\\

\hline


{\scriptsize
\KwIn{
    Non-moving node subgraph, $\mathcal{G}'=
    \{\mathcal{V}',\mathcal{E}'\}$; active moving node set \AM;
    initial graph $\mathcal{G}_{i}$
  }
\SetKwFunction{AT}{AggregateTraffic}
\SetKwFunction{AN}{AttachNode}
\SetKwFunction{GAP}{GetAllPermutations}
\SetKwFunction{RF}{RemainingFlows}

\RF{$\mathcal{G},\mathbf{p},k$} $\overset{\Delta}{=}
\sum_{j=k}^{|\mathbf{\AM }|} \sum_{v \in \mathcal{V}} \left( f_{\mathbf{p}[j],v} +
  f_{\mathbf{v,p}[j]} \right)$\;

$T_{\min}$ = \AT{$\mathcal{G}_{i}$}\;
$\mathcal{P}$ = \GAP(\AM )\;

\ForEach{$\mathbf{p} \in \mathcal{P}$}{

\ForEach{$u \in \mathcal{G}'$}{

$\mathcal{G}'_{1}$ = \AN{$\mathbf{p}[1],u,\mathcal{G}'$}\;
 $T$ = \AT{$\mathcal{G}'_{1}$}\;

\ForEach{$u \in \mathcal{G}'_1$}{
 $B_L =  T + $\RF{$\mathcal{G}'_1,\mathbf{p}, 2$}\;

\eIf{
  $(\underset{\mathcal{G}_{i} \rightarrow
    \mathcal{G}'_{1}}{d}(\mathbf{p}[1])
  \leq h_{\mathbf{p}[1]}) \&\&  (B_L < T_{\min}) $}
{
  $\mathcal{G}'_{2}$ = \AN{$\mathbf{p}[2],u,\mathcal{G}'_{1}$}\;
  $T$ = \AT{$\mathcal{G}'_{2}$}\;
  \ForEach{$u \in \mathcal{G}'_{2}$}{
 $B_L =  T + $\RF{$\mathcal{G}'_2,\mathbf{p}, 3$}\;
\eIf{$(\underset{\mathcal{G}_{i} \rightarrow \mathcal{G}'_{2}}{d}(\mathbf{p}[2])
  \leq h_{\mathbf{p}[2]}) \&\&  (B_L< T_{\min}) $}
{
  $\mathcal{G}'_{3}$ = \AN{$\mathbf{p}[3],u,\mathcal{G}'_{2}$}\;
  $T$ = \AT{$\mathcal{G}'_{3}$}\;
  \If{$(\underset{\mathcal{G}_{i} \rightarrow
\mathcal{G}'_{3}}{d}(\mathbf{p}[3]) \leq h_{\mathbf{p}[3]}) \&\& (T <
T_{\min}$)} {
  $T_{\min} = T$\;
  \AM.anchorlabel = GetAnchorLabel()\;
  \AM.desiredlabel =
  GetDesiredLabel()\;
}
}{break\;}} }{break\;}

}

}

}

}
\\
\hline

\end{tabular}

\end{algorithm}

After an optimal solution or optimal topology is obtained by using
branch-and-bound method, in order to obtain the complete desired
topology $\mathcal{G}_{f}$, the passive moving nodes have to be
attached to the optimal topology $\mathcal{G}$ that gives the minimum
aggregate traffic. Each passive moving node is attached to the optimal
topology obtained from the branch-and-bound method in such a way that
the amount of movement of the passive moving nodes is minimized,
\begin{equation}
\label{eq:min_3}
\begin{aligned}
&\underset{\mathcal{G}_{f}}{\text{minimize}} &&\sum_{v \in \PM} \underset{\mathcal{G}_{i} \rightarrow \mathcal{G}_{f}}{d}(v) \\
&\text{subject to} &&\mathcal{G} \subset \mathcal{G}_{f}.\\
\end{aligned}
\end{equation}
This can be done in a simple
iterative process, which is summarized in Algorithm~4.
\begin{algorithm}[t]

\begin{tabular}[c]{p{8cm}}

  \hline
  {\scriptsize
    \begin{bf}Algorithm 4: \end{bf} Reposition Passive
    Moving Nodes }\\
  \hline
\SetKwFunction{AN}{AttachNode}
\SetKwFunction{GAL}{GetAnchorLabel}
\SetKwFunction{GDL}{GetDesiredLabel}

{\scriptsize
  \KwIn{$\mathcal{G} = (\mathcal{V},\mathcal{E})$, graph
    with all active moving nodes attached to graph of non-moving
    nodes, $\mathcal{G}'$, according to optimization
    routine; \PM , set of passive moving nodes }

\KwOut{$\mathcal{G}_f= (\mathcal{V}_f, \mathcal{E}_f)$, final graph
  topology with passive moving nodes attached}

$\mathcal{G}_{f} = \mathcal{G}$\; 

\ForEach{$u \in \PM$} {

  $d_{\min} = \infty$ \;

\ForEach{$v \in \mathcal{V}$}{

$\mathcal{G}_{tmp}$ = \AN{u,v,$\mathcal{G}$} \;

$d = \underset{\mathcal{G}_{i} \rightarrow \mathcal{G}_{tmp}}{d}(v)$ \;

\If{$ \left(  d \leq h_{v} \right) \cap \left( d < d_{\min} \right)  $}{
  $v_{\min} =v$ \;
  $d_{\min} = d$ \;

 }

 }
$\mathcal{G}_{f}$ = \AN{u,$v_{\min}$,$\mathcal{G}_{f}$}\;

}

}
\\
\hline

\end{tabular}

\end{algorithm}

\subsection{Greedy Algorithm}
\label{sec:ga}

Even with the use of branch-and-bound to reduce the number of
solutions that must be evaluated, the complexity of finding the
optimal solution can still be very high.  This motivates us to
consider a strategy that can find a suboptimal solution to the
optimization problem but with much lower complexity.  Greedy
algorithms are strategies to address optimization problems built under
the premise that a globally optimal, or at least a good solution, can
be found by making a series of locally optimal
choices~\cite{CLRbook,Gutin}.  The greedy method is applied to find a
solution to \eqref{eq:rooted_optimization} after a root has been
determined and the root obtains the active and passive moving node
sets by using Algorithm 2.

The idea is to build an achievable tree with low aggregate traffic,
starting from a subgraph $\mathcal{G}'$ of non-moving nodes.  The
greedy algorithm is performed iteratively.  Before the first step the
working graph $\mathcal{G}$ is set equal to $\mathcal{G}'$.  At each
iterative step, every node that has not yet been assigned a position
in the working graph is evaluated.  For each such node, the achievable
locations are found for it that minimize the aggregate traffic, by
exhaustive search.

The node and its location that achieves the minimum aggregate traffic
can be formalized mathematically in the solution to
\begin{equation}
\label{eq:min_2}
\begin{aligned}
&\arg \min_{\mathcal{G}^{+}}&&\sum_{u \in \mathcal{V}_{\mathcal{G}}}\sum_{v \in
\mathcal{V}_{\mathcal{G}}}f_{uv}d_{\mathcal{G}}(u,v)\\
&\text{subject to} &&\mathcal{G}' \subset \mathcal{G}, |\mathcal{V}_{\mathcal{G}} \setminus \mathcal{V}_{\mathcal{G}'}| = 1\\
&&&y = (\mathcal{V}_{\mathcal{G}} \setminus
\mathcal{V}_{\mathcal{G}'}) \ni y \in \AM  \\
&&&\underset{\mathcal{G}_{i} \rightarrow \mathcal{G}}{d}(y) \leq h_{y}.\\
\end{aligned}
\end{equation}
The greedy algorithm is summarized in Algorithm~5.

\begin{algorithm}[t]

\begin{tabular}[c]{p{8cm}}

\hline
{\scriptsize \begin{bf}Algorithm 5: \end{bf} Greedy Algorithm }\\
\hline

 {\scriptsize
\KwIn{
    Non-moving node subgraph, $\mathcal{G}'=
    \{\mathcal{V}',\mathcal{E}'\}$; active moving node set \AM;
    initial graph $\mathcal{G}_{i}$
  }

\SetKwFunction{AN}{AttachNode}
\SetKwFunction{AT}{AggregateTraffic}

$\mathcal{G}$ = $\mathcal{G}'$\;

\While{$\AM  \neq \emptyset$}{
  $T_{\min}=\infty$\;
  \ForEach{$u \in \AM$} {

\ForEach{$v \in \mathcal{V}_{G}$}{

$\mathcal{G}_{tmp}$ = \AN{$u,v,\mathcal{G}$}\;
$T=$ \AT{$\mathcal{G}_{tmp}$}\;

\If{$ (\underset{\mathcal{G}_{i} \rightarrow \mathcal{G}_{tmp}}{d}(v)
  \leq h_{v}) \&\& (T < T_{\min})  $}
{
  $T_{\min}=T$\;
  $u_{\min}=u$\;
  $v_{\min}=v$\;
}

 }

 }

$\mathcal{G}$ = \AN{$u_{\min},v_{\min},\mathcal{G}$}\;
 $\AM = \AM \cap \overline{v_{\min}}$\;

}

}
\\
\hline

\end{tabular}

\end{algorithm}

After the active nodes are assigned positions, the root then assigns
the positions of each node in the passive moving node set, using
Algorithm~4.



\section{Complexity Analysis}
\label{sec:complex}

In this section, we briefly analyze the complexity of the optimization
algorithms given in \autoref{sec:opt}.  The complexity of the algorithms
is important because it gives a guideline as to how useful these
algorithms will be when applied to large networks with large active
moving sets.

The worst case running time for the optimal algorithm occurs under the
following conditions:
\begin{itemize}
\item all of the nodes except the root are active moving nodes,
\item all of the nodes have a very large maximum possible hop,
  $h_{v}$, such that they can be moved to any part of the graph, and
\item the value of each solution found from the solution tree is
  monotonically decreasing.
\end{itemize}
The last condition requires that the entire solution tree has to be
traversed.  Let $n$ denote the number of nodes in $\mathbf{G}_i$.
Then the worst case running time for the optimal algorithm is
\begin{equation}
\label{eq:time_2} T(n) = ((n - 1)!)^2
\end{equation}
Hence $T(n) \in O((n!)^2) \le O(n^{2n})$.  Thus, although the optimal
algorithm could be used to obtain the graph topology to
achieve the minimum aggregate traffic, this algorithm
has high complexity when the size of the active moving set becomes large.

The worst case running time for the greedy algorithm also occurs when all the
nodes except the root are active moving nodes  and is given by
\begin{equation}
\label{eq:time_1} T(n) = \sum_{i = 1}^{n -1 }i(n-i),
\end{equation}
It is easy to show that $T(n) \in O(n^3)$, and the greedy algorithm has
polynomial-time complexity.  Thus, the greedy algorithm has a much
lower complexity for large networks than the optimal algorithm;
however, it is not guaranteed to find the optimal solution.



\section{Simulation Results}
\label{sec:sim_result}

In this section, we evaluate and compare the performance of both the
greedy and exact algorithms to minimize aggregate traffic for small
networks of 3 to 15 nodes. For each size network, we distribute a
total flow of 1~Mbps randomly among all possible source-destination
pairs according to a uniform distribution.  A total of 50 different
flow allocations are used to generate our numerical results.  For each
flow allocation, an initial tree topology is randomly selected from
among the possible trees for the network.  Finally, for each topology
and data flow, the maximum possible hop $h_{v}$ at each node is
selected randomly according to a uniform random variable ranging from
0 to a specified $h_{\max}$.  We repeat the same experiments for
different values of $h_{\max}$.  We report results for $h_{\max}$
values of 1, 3, and 10 hops.  For the optimal algorithm, we report
results for network sizes up to 7 nodes.  Beyond that, the complexity
of the optimal algorithm required too much running time.

\begin{figure} 
\centering
 \includegraphics[width=0.48\textwidth]{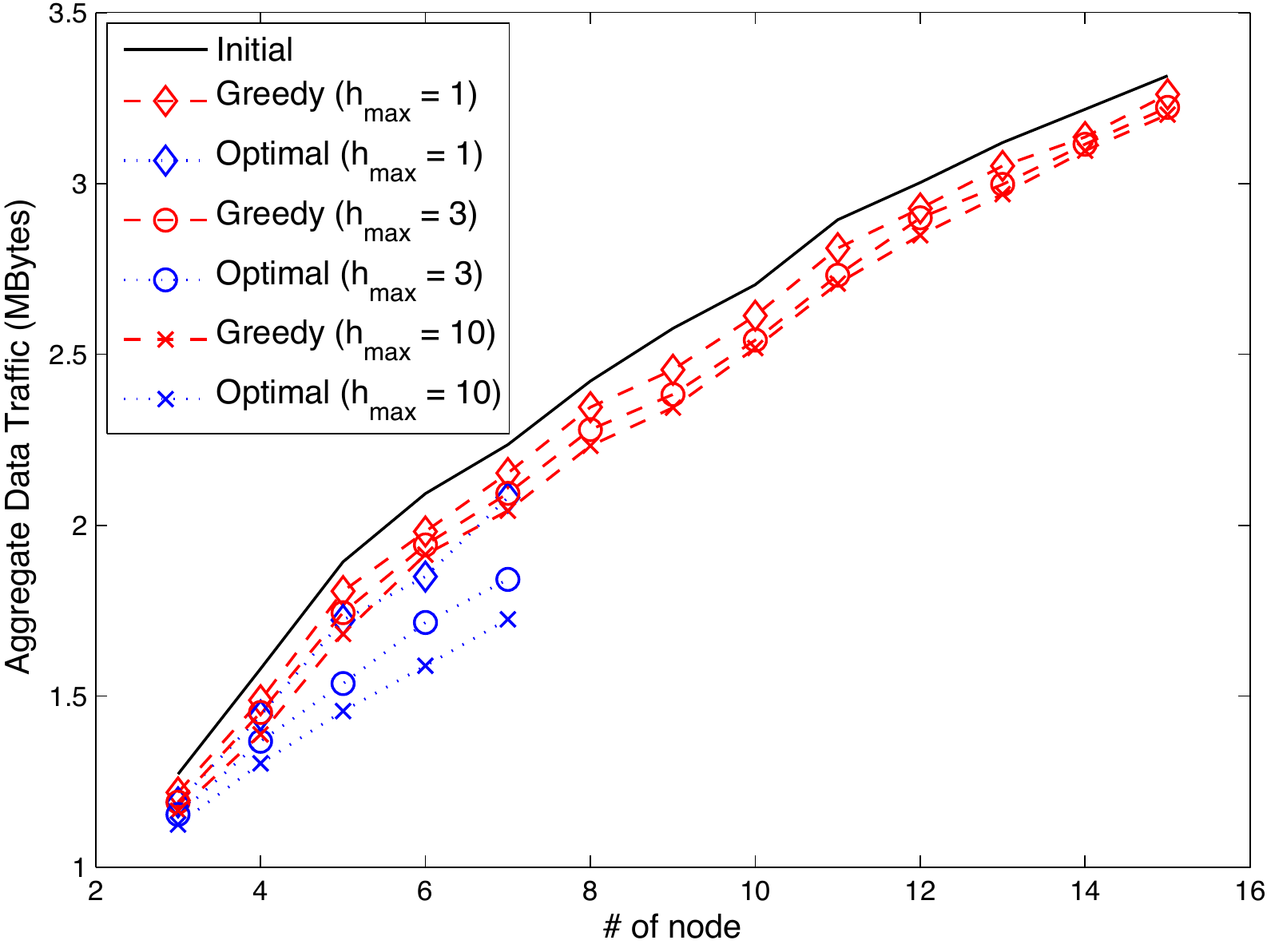}
 \caption{Minimum achievable aggregate traffic for greedy and optimal
   algorithms as a function of network size.}
 \label{fig:result_1}
\end{figure}

The average aggregate traffic achieved by the optimal and greedy
algorithms is shown in \autoref{fig:result_1} as a function of the
network size.  Also shown is the initial aggregate traffic before
optimization.  It can be observed that when the network size is small,
the greedy and optimal algorithms provide similar performance, since
there are a limited number of possible candidate topologies
to be considered. A root does not have many options to reposition its own
children in $\mathcal{G}_{i}$, and the nodes in the small network are
already close to each other.   Thus, little reduction in aggregate
traffic is possible.

As the network size grows larger, the amount of energy at each node
that is available for repositioning plays an important role in the
final aggregate traffic, especially for the optimal algorithm.  For
instance, for a network with seven nodes, if $h_{\max}=1$, the minimum
aggregate traffic is approximately 2.  If $h_{\max}=10$, the minimum
aggregate traffic is approximately 1.6.  Thus, the optimal algorithm
is able to leverage the additional degrees of freedom to better
reconfigure the network.  On the other hand, large $h_{\max}$ also
translates into more feasible network topologies, which can slow the
execution of the optimization algorithm.  The greedy algorithm does gain
from increasing the amount of allowed movement, but not as dramatically as the
optimal algorithm.

\begin{figure} 
\centering
 \includegraphics[width=0.5\textwidth]{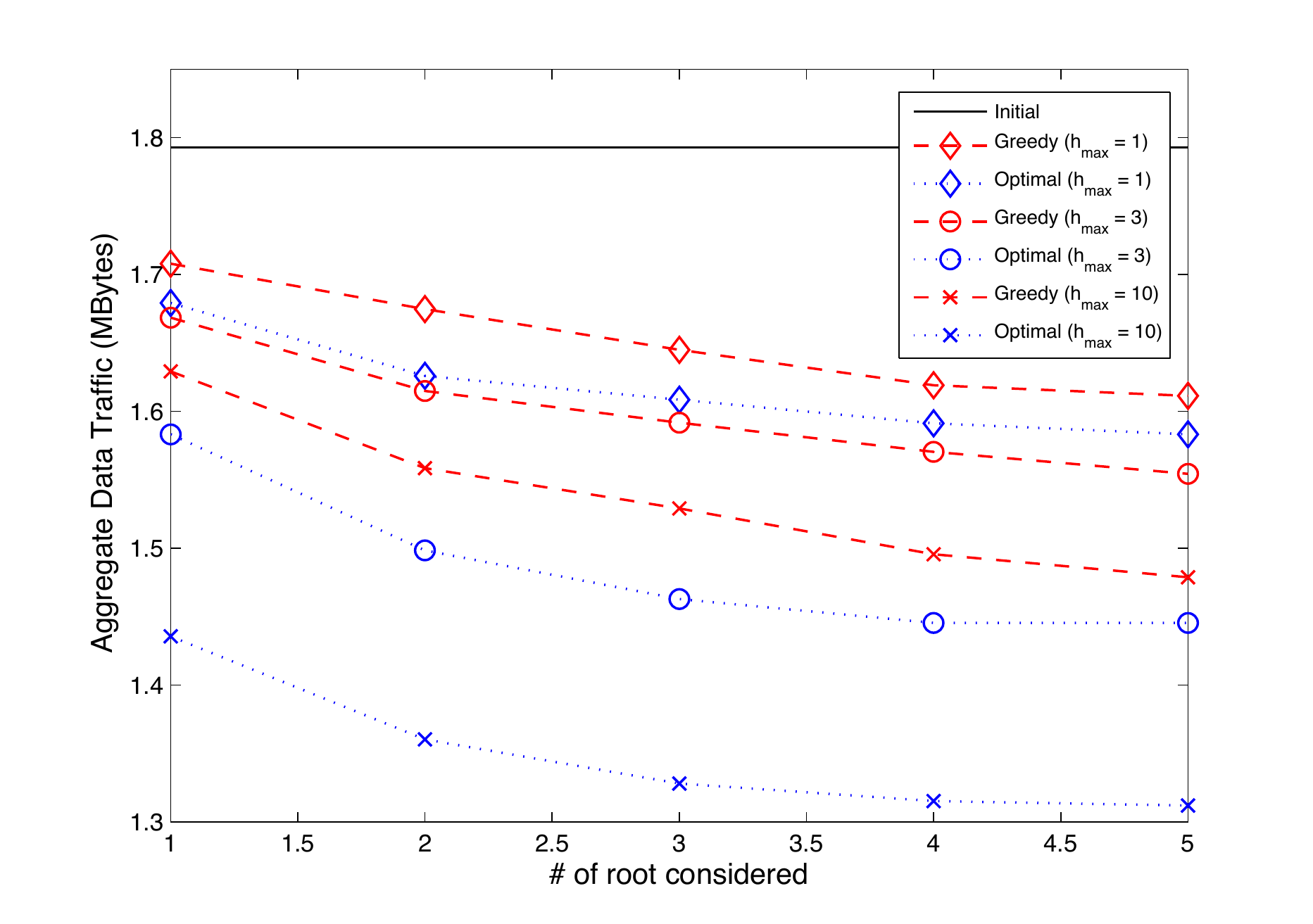}
 \caption{Minimum achievable aggregate traffic for greedy and exact
   algorithms with multiple root selection.}
 \label{fig:result_2}
\end{figure}

The performance of both algorithms can potentially be improved by
considering the best aggregate traffic that can be achieved for
multiple choices of the tree's root.  We fix the network to consist of
five nodes, and we choose the flows and topologies randomly as before.
For each topology, we select multiple roots at random and choose the
root that results in the minimum aggregate traffic. The results are
averaged over all generated random topologies with different flow
allocations. The results are shown in~\autoref{fig:result_2} for
different values of $h_{\max}$ as a function of the number of roots
considered.

It can be observed that the selection of the root affects the minimum
aggregate traffic that can be achieved. The more roots that are
considered, the lower the minimized aggregate data traffic for both
the greedy and optimal algorithms.  Similarly, the higher $h_{max}$,
the lower minimized aggregate traffic since there will be more
candidate solutions.  The selection of a root matters since each node
in the network may have different limited amount of movement. If each
node in the network has an unlimited amount of movement, the network
topology can be transformed to any network topology using the method
given in~\autoref{sec:network}, no matter which node is selected to be
a root.  However, some node that is selected to be a root in
$\mathcal{G}_{i}$ may lead to more achievable network topologies than
when others are selected as the root. Hence the root selection is one
of an important issue for the future work.

\section{Conclusion}
\label{sec:con}

In this paper, we developed algorithms to reconfigure the network
topology of a systems of mobile robots to minimize the aggregate
traffic in the network, under a constraint on the amount of energy
available for movement by each robot.  We also constrain our network
to maintain network connectivity at all times, and so we develop our
optimization algorithms under a framework in which the robots are
routed through the network in such a way that network connectivity is
maintained.  We developed optimal and greedy algorithms to minimize
the aggregate traffic under the specified constraints, and we provide
complexity and performance comparisons.  The results show that
although both algorithms can decrease the aggregate traffic, the
greedy algorithm does not achieve performance close to that of the
optimal algorithm.  On the other hand, the greedy algorithm has only
polynomial complexity, versus factorial-squared complexity for the
optimal algorithm.  The results also show that the performance of both
algorithms improve with the amount of energy available for node
movement and with the number of different roots for which the
aggregate traffic is evaluated.

\bibliographystyle{IEEEtran}

\end{document}